\documentclass[pdflatex,sn-nature]{sn-jnl}

\usepackage{lineno}
\usepackage{graphicx}%
\usepackage{multirow}%
\usepackage{amsmath,amssymb,amsfonts}%
\usepackage{amsthm}%
\usepackage{mathrsfs}%
\usepackage[title]{appendix}%
\usepackage{xcolor}%
\usepackage{textcomp}%
\usepackage{manyfoot}%
\usepackage{booktabs}%
\usepackage{algorithm}%
\usepackage{algorithmicx}%
\usepackage{algpseudocode}%
\usepackage{listings}%
\usepackage{textcomp}%

\begin{document}

\title[Article Title]{Topological Solitons in Square-root Graphene Nanoribbons Controlled by Electric Fields}


\author[1]{\fnm{Haiyue} \sur{Huang}}\email{haiyuehuang@g.ucla.edu}
\equalcont{These authors contributed equally to this work.}

\author[2]{\fnm{Mamun} \sur{Sarker}}\email{msarker2@unl.edu}
\equalcont{These authors contributed equally to this work.}

\author[3]{\fnm{Percy} \sur{Zahl}}\email{pzahl@bnl.gov}

\author[4]{\fnm{C. Stephen} \sur{Hellberg}}\email{carl.s.hellberg.civ@us.navy.mil}
\author[5]{\fnm{Jeremy} \sur{Levy}}\email{jlevy@pitt.edu}
\author*[1]{\fnm{Ioannis} \sur{Petrides}}\email{ipetrides@g.ucla.edu}
\author*[2]{\fnm{Alexander} \sur{Sinitskii}}\email{sinitskii@unl.edu}
\author*[1,6]{\fnm{Prineha} \sur{Narang}}\email{prineha@ucla.edu}

\affil[1]{\orgdiv{Division of Physical Sciences, College of Letters and Science}, \orgname{University of California}, \orgaddress{\city{Los Angeles}, \postcode{90095}, \state{California}, \country{USA}}}

\affil[2]{\orgdiv{Department of Chemistry}, \orgname{University of Nebraska}, \orgaddress{\city{Lincoln}, \postcode{68588}, \state{Nebraska}, \country{USA}}}

\affil[3]{\orgdiv{Center for Functional Nanomaterials}, \orgname{Brookhaven National Laboratory}, \orgaddress{\city{Upton}, \postcode{11973}, \state{New York}, \country{USA}}}

\affil[4]{\orgname{U.S. Naval Research Laboratory}, \orgaddress{\city{Washington}, \postcode{20375}, \state{District of Columbia}, \country{USA}}}

\affil[5]{\orgdiv{Department of Physics and Astronomy}, \orgname{University of Pittsburgh}, \orgaddress{ \city{Pittsburgh}, \postcode{15260}, \state{Pennsylvania}, \country{USA}}}

\affil[6]{\orgdiv{Department of Electrical and Computer Engineering}, \orgname{University of California}, \orgaddress{\city{Los Angeles}, \postcode{90095}, \state{California}, \country{USA}}}


\abstract{Graphene nanoribbons (GNRs) are unique quasi-one-dimensional (1D) materials that have garnered a lot of research interest in the field of topological insulators. While the topological phases exhibited by GNRs are primarily governed by their chemical structures, the ability to externally control these phases is crucial for their potential utilization in quantum electronics and spintronics. Here we propose a class of GNRs featured by mirror symmetry and four zigzag segments in a unit cell that has unique topological properties induced and controlled by an externally applied electric field. Their band structures manifest two finite gaps which support topological solitons, as described by an effective square-root model. To demonstrate the experimental feasibility, we design and synthesize a representative partially zigzag chevron-type GNR (pzc-GNR) with the desired zigzag segments using a bottom-up approach. First-principles calculations on pzc-GNR reveal band inversions at the two finite gaps by switching the direction of the electric field, which is in accordance with predictions from the square-root Hamiltonian. We show different topological phases can be achieved by controlling the direction of the field and the chemical potential of the system in square-root GNRs. Consequently, upon adding a step-function electric field, solitons states can be generated at the domain wall. We discuss the properties of two types of soliton states, depending on whether the terminating commensurate unit cell is mirror symmetric.}

\maketitle

\section{Main}\label{sec1}

Graphene nanoribbons (GNRs) can be regarded as quasi-one-dimensional materials cut from graphene or unrolled from carbon nanotubes \cite{ezawa2006peculiar,PhysRevB.73.235411}. Advancements in bottom-up synthesis methods have led to a growing variety of GNRs with unique edge structures, such as zigzag, armchair, chevron, and chiral edged \cite{cai2010atomically,ruffieux2016surface,huang2023sub,liu2020chevron,dobner2022diffusion}. Among these, GNRs exhibiting topological phases \cite{rizzo2018topological, li2021topological, lee2018topological,jiang2020topology} have obtained significant attention as they serve as a versatile platform for the exploration of fundamental topological models \cite{cao2017topological,lin2018topological}, including the well-known Su-Schrieffer-Heeger (SSH) model \cite{groning2018engineering}. The bulk-boundary correspondence \cite{vanderbilt2018berry} ensures that nontrivial topological GNRs can exhibit a soliton-like localized state within the band gap when a boundary is made. These zero-dimensional boundary states derived from the quasi-1D bulk are expected to be a topological analogue of quantum dots \cite{rizzo2021rationally}, which are promising candidates for carbon-based quantum electronics and spintronics.



To engineer the topological phases of GNRs, there are in general two strategies. The first strategy is to control the chemical structures of GNRs by designing the reactions during the bottom-up synthesis. In this case, the boundary state is formed either at a structural heterojunction \cite{jiang2020topology,rizzo2018topological,cao2017topological} or at the two ends \cite{li2021topological}. The second strategy also has some requirements on the chemical structures but at the same time involves an externally controllable factor to manipulate the electronic structure and the topological phase. As the latter is externally controlled, the as-formed boundary states are more versatile, and can be better integrated in electronic devices. So far, the realization of the second strategy for GNRs has been demonstrated in only a few examples. \cite{zhao2021topological,huang2024strain}. These studies are still in their early stages and have not started to address experimental feasibility.

In this work, we seek to broaden the scope where the band structure manipulation and topological phase transition can be achieved externally even after the synthesis is finished. To this end, we introduce a whole class of GNRs of which topological phases are induced and controlled by a transverse electric field. The common feature of GNRs in this class is that their topological properties can be explained and predicted by a Hamiltonian belonging to the general category of the ``square-root" model \cite{kremer2020square,arkinstall2017topological,kane2014topological}. Taking the square root of a Hamiltonian has been shown to generate novel nontrivial topological phases. For example, taking the square root of the Hamiltonian of two coupled Rice-Mele chains yields a new and nontrivial model, at the expense of broken crystal symmetries. The new model possesses mirror symmetry and nonsymmorphic chiral symmetry, which leads to a 1D topological crystalline insulator \cite{fu2011topological}. 
To demonstrate topological phase transition and generation of soliton states by electric fields in this class of GNRs, we focus our calculations on a representative GNR with partially zigzag chevron type edges (pzc-GNR). Additionally, we provide evidence on the synthesis and characterization of a representative GNR with partially zigzag chevron type edges (pzc-GNR), which demonstrates the experimental accessibility of the design. 

\section{Design of square-root GNRs}\label{sec2}
The band structures of GNRs are known to depend strongly on the width and edge structures \cite{ezawa2006peculiar, wakabayashi2010electronic}. Different from armchair edges, zigzag edges lead to localized states with energies near the Fermi level. Lining up zigzag-like segments in a strategic way can therefore form bands with dispersion and modulate the density of states (DOS) of graphene near the Fermi level. To design square-root GNRs, we place four zigzag segments in a mirror-symmetric way (Fig. 1a and 1b). The hopping between zigzag segments is effectively simplified to alternate between two distinct values (\(t_{1}\) and \(t_{2}\)). When a transverse electric field is applied, the electrostatic potential of electrons is changed, including those near the zigzag edges. 
For an in-plane electric field, the two top (bottom) zigzag segments have an increase (decrease) of their potential by \emph{m}, see Fig. 1c. In this regime, GNRs near the Fermi level define a quasi-1D tight-binding model where the unit cell contains four lattice degrees of freedom with on-site potentials and coupled by nearest-neighbor hopping, see Fig. 1d. The Hamiltonian of this effective model in the momentum space is described by:

\begin{equation}
H(k)= \begin{pmatrix}
 m&t_{2}e^{-isk}&0&t_{1}e^{i(\frac{b}{2}-s)k} \\
 t_{2}e^{isk}&-m&t_{1}e^{-i(\frac{b}{2}-s)k}&0 \\
 0&t_{1}e^{i(\frac{b}{2}-s)k}&-m&t_{2}e^{-isk} \\
 t_{1}e^{-i(\frac{b}{2}-s)k}&0&t_{2}e^{isk}&m
 \end{pmatrix}
\end{equation}
where \emph{m} and \emph{–m} are the on-site potentials determined by the electric field, \(t_{1}\) and \(t_{2}\) are hopping parameters, as indicated by dash and solid lines, \emph{s} is the distance between the first and second lattice sites, \emph{k} is the wave vector, and \emph{b} is the length of the unit cell. Because the model has a mirror symmetry, the distance between the second and third sites is (\(\frac{b}{2}-s\)). 
This effective Hamiltonian has a square-root origin and presents nontrivial topological features as we will discuss later. There are numerous possible structures of GNRs where the above procedure results in a square-root Hamiltonian. The multitude is further expanded when considering that the zigzag segment is only conceptual. Chemically, it can contain a mixture of more than one zigzag edges with other edge configurations, such as armchair and chevron. In this study, we use a representative pzc-GNR to demonstrate our primary findings. Further information regarding the tight-binding model and another example GNR can be found in Supplemental Materials. As the GNRs considered here share the same set of topological properties, we dub them ``square-root GNRs".

 To synthesize pzc-GNRs, we deposit 2,7-dibromo-10,13-dimethyl-9,14-diphenylbenzo[f]tetraphene (monomer-T) onto an Au(111) surface by sublimation in an ultra-high vacuum (UHV) (Fig. 1e, top). The molecules are then coupled using Ullmann coupling by annealing at 280°C for 10 minutes, forming long-chain homo-polymers (polymer-T in Fig. 1e, middle). These polymers are transformed into pzc-GNRs through a high-temperature on-surface cyclodehydrogenation reaction. Most of the polymers transform during the second step annealing process at 380°C for 5 minutes, and a complete transformation occurs during the final, third step of annealing at 450°C for 2 minutes, where the polymers undergo cyclization to form the desired GNRs (Fig. 1e bottom).

Fig. 1f displays a representative large-area topographic image recorded by scanning tunneling microscopy (STM) on a polymer-T sample after it is annealed at 280°C. In this STM image, the edges of the polymer chains demonstrate bright dots, which correspond to the nonplanar uncyclized fragments of the polymer. Importantly, these bright features are not regular and the chains also contain extended darker regions, suggesting that at this temperature the polymer-T is already partially cyclized. The large-area STM feedback scan taken on a sample after annealing at 380°C shows that most of the bright features at the edges of the polymer chains disappear (Fig. 1g), although some rare bright dots can still be observed. A low-coverage STM image recorded on a sample after annealing at 450°C (Fig. 1h) demonstrates a complete cyclodehydrogenation reaction of the polymer chains, where all bright dots completely disappear. This process results in the formation of the desired planar pzc-GNRs. Fig. 1i shows a close-up topographic STM image of an isolated pzc-GNR on Au(111) after annealing at 450°C. The chemical structure overlay on this STM image exhibits the formation of the desired ribbon. A high-resolution non-contact atomic force microscopy (nc-AFM) image shown in Fig. 1j demonstrates a perfect agreement of the structure of pzc-GNR with the chemical structure shown in Fig. 1a.

\section{Band structure calculations}\label{sec3}
The band structure of the pzc-GNR is obtained by density-functional theory (DFT) calculations performed with the Quantum Espresso package. To emphasize the change upon applying the electric field, we plot the band structure of pzc-GNR under a static field with positive (+0.1V/\AA) and negative strength (–0.1V/\AA) in Fig. 2a and 2e, respectively, with comparison to the situation with zero fields (middle four bands plot in blue dashed line, other bands are omitted for clarity). It can be seen that the central gap is not altered much by the applied electric field. Importantly, the field opens up two gaps at \(k=\pi/a\). These two new gaps (0.20 eV) are denoted as finite gaps, to differentiate them from the central gap. 
The effective Hamiltonian describing the middle four bands, originating from the zigzag segments of GNR, is solved and the band structure is plotted in Fig. 2c and 2g. The set of parameters (\(t_{1}\), \(t_{2}\), and \emph{m}) is chosen such that the gaps width match that of DFT calculations (see details in Supplemental Material, Sec. 3). The two band structures calculated from DFT and the effective Hamiltonian agree with each other.

To characterize the topology of the quasi-1D system, we use the Zak phase (\(\gamma_{n}\)) \cite{resta2000manifestations}, a quantity a-priori gauge-dependent. For a mirror-symmetric system, the gauge ambiguity is lifted when the origin of the coordinate axis is set to the mirror plane, hence the Zak phase only takes 0 or \(\pi\) (mod \(2\pi\)). In this case, the Zak phase is directly related to the bulk electronic polarization by 
\begin{equation}
P=\frac{1}{2\pi}\sum_{n}^{occ}\gamma_{n}
\end{equation}
where \emph{n} is the band index, and the sum is over occupied bands \cite{WOS:000249021500002}. A quantized nonzero (\(\pi\)) total Zak phase indicates an nonzero polarization, and thus an nontrivial phase. The Zak phase for each of the four bands can be calculated according to  
\begin{equation}
    \gamma_{n}= i \oint_{BZ} dk \langle u_{n,k}| \partial_{k}|u_{n,k} \rangle
\end{equation}
where \(u_{n,k}\) is the \emph{n}th periodic part of the Bloch function \(\psi_{n,k}\) and BZ is the Brillouin zone in one dimension. With mirror symmetry, the calculation can be simplified to
\begin{equation}
    e^{i\gamma_{n}}=\langle \psi_{n,\Gamma}|M|\psi_{n,\Gamma}\rangle \langle\psi_{n,X}|M|\psi_{n,X}\rangle
\end{equation}
where \emph{M} is the mirror symmetry operator, \(\mathit{\Gamma}\) means \(k=0\), and \emph{X} is the middle of the BZ with \(k=\pi/a\).

Eq. 4 relates the parity of the wave function to the Zak phase of the band. For our system, the lower pair of bands switch parity at the \emph{X} point under opposite directions of electric field, and so does the upper pair, see Fig. 2b and 2f. The parity at the gamma point, however, persists. This indicates a band inversion at the two finite gaps caused by flipping the direction of the electric field. The description of the system using the effective Hamiltonian (Eq. 1) is shown in Fig. 2c and 2g, with the wave function weights illustrated in Fig. 2d and 2h. The parity at \emph{X} matches well with that in Fig. 2b and 2f, suggesting the band inversion behavior of GNR has a fundamental origin described by our effective Hamiltonian.

The polarization, i.e., total Zak phase at half filling remains zero and independent of the direction of the field applied. In contrast, when the system is gated such that a typical highest occupied band (HOMO) becomes unoccupied, or the lowest unoccupied band (LUMO) becomes occupied, the total Zak phase is determined by the direction of the electric field. This feature is not influenced by the higher- or lower-energy bulk bands, highlighting the robustness of the topological phase for square-root GNRs in response to an electric field.

\section{Topological phase diagram}\label{sec4}
To reveal the total Zak phase in pzc-GNR, we 
calculate the location of Wannier charge centers (\(\bar{r}_{n}\)) \cite{Pizzi2020} from DFT results, and determine the polarization according to:
\begin{equation}
    P=\frac{1}{a}\sum_{n}^{occ}\bar{r}_{n}
\end{equation}
As pzc-GNR has 154 valence bands, we are interested in the case in which \emph{n} is iterated from 1 to 153, and from 1 to 155. Utilizing the WANNIER90 package, the Wannier centers of the lower 153 and 155 bands are obtained, which then gives the polarization according to Eq. 5. Using polarization as the topological invariant, we can then obtain the phase diagrams shown in Fig. 3a, 3b, and 3c. With 153 occupied bands, a quantized polarization of 0.5 is found for the positive field, which goes to 0 when the direction of the field is flipped. A similar quantized behavior but with an opposite trend is observed when the occupation is 155.

A straightforward way to understand the phase transition can be obtained by considering the four-band square-root Hamiltonian. 
Flipping the direction of the electric field effectively shifts the Wannier center(s) along the y axis. With only one occupied band, the Wannier center can only be located either at the boundary of the cell (up) or at the center of the cell (down) due to mirror symmetry. Therefore, switching the electric field and moving the Wannier centers vertically inevitably shifts their respective x coordinates. The system then alternates between the two configurations shown in Fig. 3d., which results in a switch of the total polarization between 0 and 0.5. Similar logic applies to the case with 3 occupied bands. This can also explain that square-root GNR, when not gated, does not have a tunable topological phase (Fig 3b and 3e). Upon flipping the electric field, the two Wannier centers corresponding to two occupied bands switch y coordinates, which results in the same summation of x coordinates.


\section{Topological solitons}\label{sec5}
A nonzero polarization is a bulk property ensuring the existence of robust boundary states, according to the bulk-boundary correspondence. To directly demonstrate such a topologically protected boundary state, an edge or an interface must be created. This can be achieved either by terminating the structure (joining with vacuum, which is a trivial insulator), or joining with a material in a different phase. It should be noted here that the topological phase calculated above is based on a specific choice of unit cell with mirror symmetry (Fig. 4a, inset). The first method requires the structure to have end structures commensurate with the unit cell \cite{rhim2017bulk}, which is experimentally not realistic for pzc-GNR. Instead, we can implement a spatially varying electric field to create an interface between a nontrivial and a trivial phase. The electric field is thus applied such that the domain wall is located at the edge of the mirror-symmetric unit cell (Fig. 4c). According to previous discussions, the left and right regions now belong to different topological phases, when pzc-GNR is filled up to the \(153^{rd}\) band or \(155^{th}\) band. Indeed, first-principles calculations reveal that boundary states do exist with energy levels within the finite gaps (Fig. 4d, filled peaks), and are localized at the domain wall, as expected. In comparison, if the system is under a homogeneous transverse electric field, there is no sign of such in-gap states (Fig. 4d, black and gray curves). We also simulate a finite pzc-GNR molecule and observe that the in-gap state localized at the domain wall still exists (see Supplemental Materials Sec. 5).

An alternative consideration includes another type of termination with broken mirror symmetry. This situation is simulated by applying the same profile of electric field to a supercell made of non-mirror-symmetric unit cells (Fig. 4b, right). Boundary states generated in this case can also be localized and have energy levels within the gaps (Fig. 4b and 4e). However, due to the associated underlying symmetries, boundary states shown in Fig. 4a and 4b corresponding to the two choice of unit cells, hence distinct domain wall configurations exhibit fundamentally different properties. Those from Fig. 4b are non-degenerate and their energy levels are dependent on the width of the domain wall, i.e., how abruptly the electric field alters direction (see Supplemental Materials Sec. 6). The non-mirror-symmetric unit cell results in solitons, akin to the Rice-Mele model, which is known to support non-quantized topological invariants \cite{brzezicki2020topological, allen2022nonsymmorphic, rice1982elementary}. Such Rice-Mele type of soliton states lead to nonquantized charge due to lack of mirror symmetry. The ability to form both topological and Rice-Mele solitons in square-root GNRs by adding step-function electric fields is a unique feature of our proposed system. Metallic GNRs, such as 5-sGNR, generate only Rice-Mele type of soliton upon adding similar electric fields \cite{rizzo2020inducing} (Supplemental Materials Sec. 6).

\section{Summary}\label{sec6}

In this work, we introduce a class of GNRs with topological properties controlled by a transverse electric field. This class of GNRs is mirror symmetric and has a central gap and two finite gaps. The topological features of GNRs are delineated by an effective square-root-type Hamiltonian, which shows that the two finite gaps can become nontrivial depending on the direction of the electric field, regardless of low- or high-lying bulk bands. Topological soliton states can be generated with energy levels within the finite gaps, at an interface where a change of sign of the electric field occurs. On the other hand, when the termination of the domain is not commensurate with a mirror symmetric unit cell, a Rice-Mele type of soliton is supported, originating from the nonsymmorphic chiral symmetry.

The electric-field tunable phases make it possible to externally create, move, and annihilate soliton states in GNRs. They can be regarded as mobile and tunable versions of quantum dots. Their locations and spread can be controlled by the position of the domain wall and the strength of the applied electric field. When multiple of solitons are overlapped, it will be interesting to study their ground state spin configurations, where the exchange interaction can be adjusted by tuning the effective hopping due to overlap. This is important towards GNRs for spintronics, where exchange interaction is essential for spin manipulation \cite{levy2002universal,meier2003quantum, wang2021graphene,mishra2020topological}.

\section{Methods}\label{sec7}

Experimental details about the synthesis and characterization of pzc-GNRs are listed in Supplemental Materials Sec. 1. First-principles calculations were carried out using Quantum Espresso package \cite{QE-2009}. Structures were relaxed until all forces fall below 0.025 eV/\AA. Standard solid-state pseudopotentials (SSSP[1.3.0][PBE][efficiency] \cite{prandini2018precision}), a kinetic energy cutoff of 100 Ry for wave functions, a kinetic energy cutoff of 800 Ry for charge density, 16 k-points for scf, 24 k-points for nscf were used.  20 \text{\AA} of vacuum was added in the non-periodic directions. Electric field was added in the form of a saw-tooth potential by setting tefield to true. Amplitude of wave function in Fig. 2b and 2f were obtained by tight-binding calculations implemented using the PythTB package assuming nearest-neighbour hopping of \(p_{z}\) orbital of carbon atoms. The hopping parameter was set to -2.7 eV. Calculations in Fig. 4 were implemented using the CP2K package \cite{kuhne2020cp2k}. Structures were relaxed until all forces are below 0.015 eV/\AA. Perdew-Burke-Ernzerhof (PBE) functional, double-zeta valence basis set, and pseudopotentials of Goedecker- Teter- Hutter (GTH) were used. External potential was added according to the function described in Supplemental Materials Sec. 5. To get DOS from eigenvalues, a smearing of 0.5 meV was adopted.


\bibliography{article}

\section{Acknowledgements}\label{sec8}
Authors thank helpful discussions with Dr. Zhigang Song and Dr. Xuecheng Tao. This work is supported by Office of Naval Research through Multidisciplinary University Research Initiative (No. N00014-20-S-F003). This work uses LT-STM/NC-AFM facility of the Center for Functional Nanomaterials (CFN), which is a U.S. Department of Energy Office of Science User Facility, at Brookhaven National Laboratory under Contract No. DE-SC0012704. Authors thank resources of the National Energy Research Scientific Computing Center (NERSC).

\section{Author information}
These authors contributed equally: Haiyue Huang, Mamun Sarker

\subsection{Contributions}
P.N., A.S., I.P., and J.L. conceived and supervised this work. H.H. elaborated on the initial concept and conducted calculations, aided by I.P. and C.S.H.. M.S. designed and synthesized precursor molecules and the GNR. M.S. and P.Z. characterized the as-synthesized GNR. H.H., M.S., and I.P. plotted the figures with contributions from other co-authors. H.H., M.S., and I.P. wrote the manuscript with inputs from other co-authors.

\subsection{Corresponding authors}
Correspondence to Prineha Narang
\clearpage
\begin{figure}[h]
\centering
\includegraphics[width=0.9\textwidth]{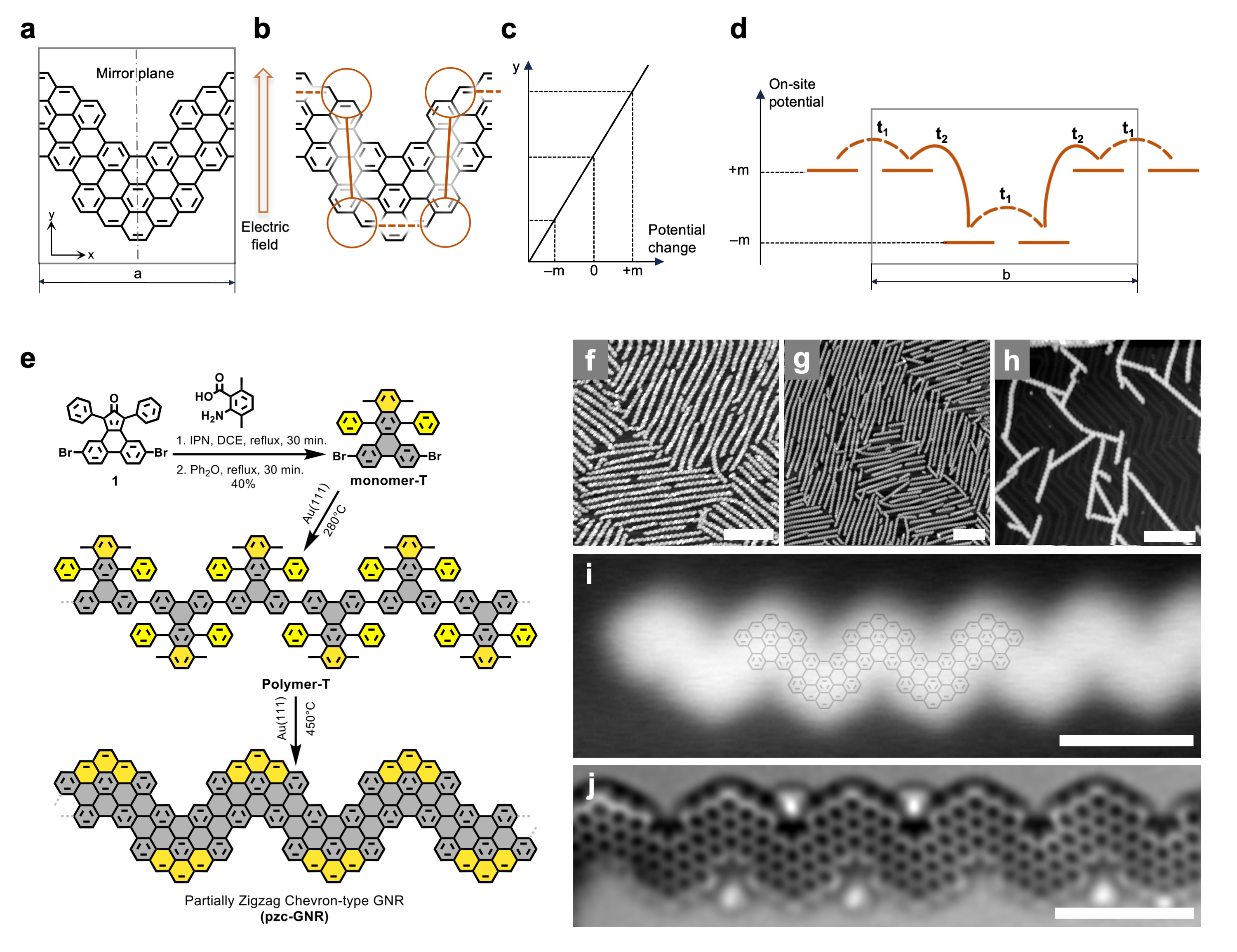}
\caption{Structure and synthesis of pzc-GNR, a representative square-root GNR. (a) The structure of a unit cell of the pzc-GNR. It has a mirror plane and contains four zigzag segments. (b) The coupling strength between those segments alternate and are illustrated by straight and dashed lines. (c) A transverse electric field is applied to the GNR and changes the electrostatic potential energy of electrons, including those in the four zigzag segments. The electronic bands near the Fermi level can be mapped to a square-root model (d), which is characterized by four lattice sites containing mirror symmetric on-site energy terms and coupled by alternating hopping strength \(t_{1}\) and \(t_{2}\). (e) Reaction scheme for the synthesis of monomer-T and pzc-GNR. (f) STM image for the polymer-T at 280°C. (g) Large-area STM scan of mostly planar pzc-GNR at 380°C. (h) STM image for completely planar pzc-GNR at 450°C. (i) Close-up STM image of an isolated pzc-GNR at 450°C. (j) High-resolution nc-AFM image of pzc-GNR. Scale bars: f-h = 20 nm, i-j = 2 nm.}\label{fig1}
\end{figure}

\clearpage

\begin{figure}[h]
\centering
\includegraphics[width=0.9\textwidth]{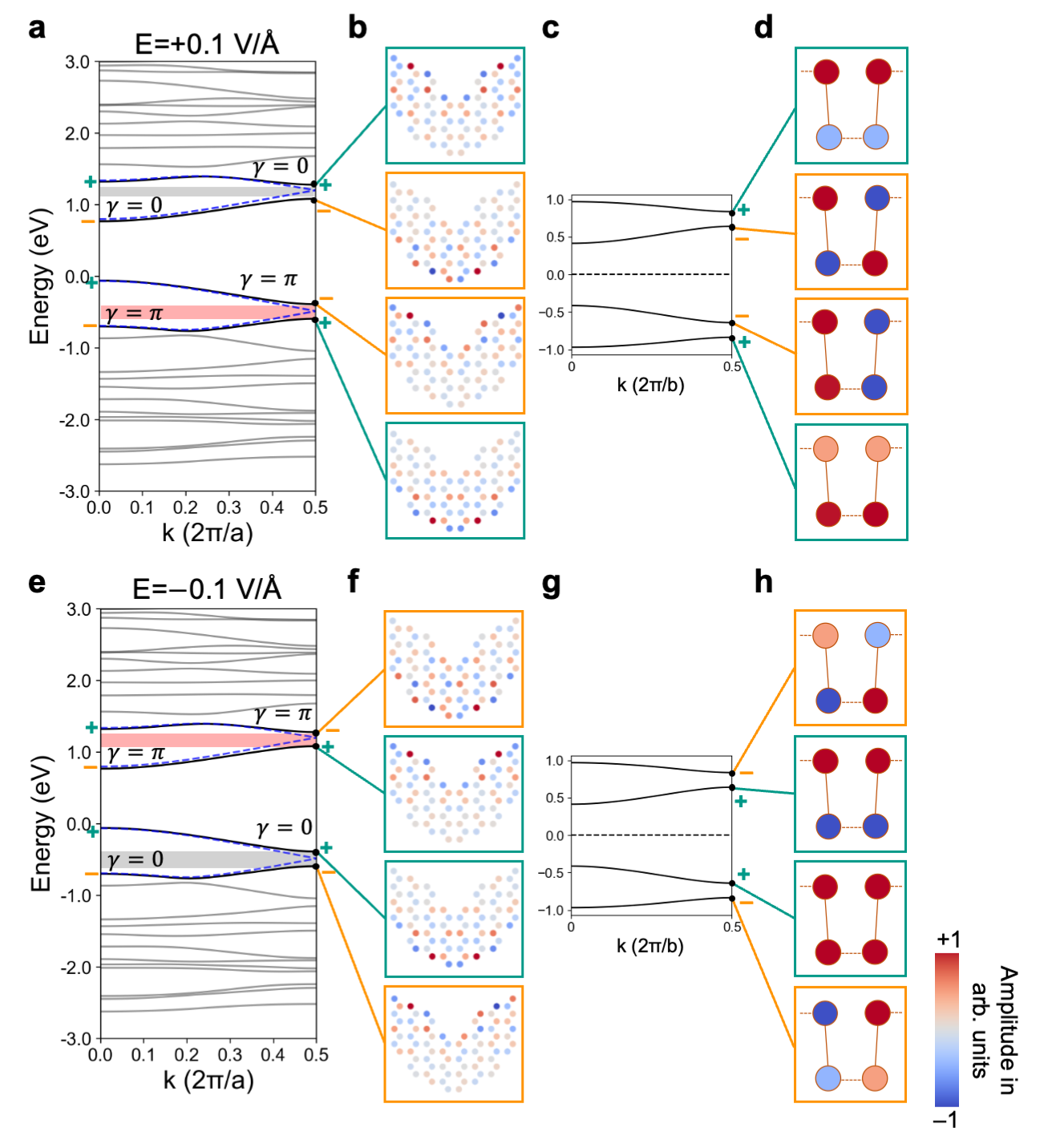}
\caption{Electronic band structures of pzc-GNR and band inversions by an electric field. (a) and (e) are band structures of pzc-GNR under a +0.1V/\AA, and a -0.1V/\AA\ electric field, respectively. The middle four bands (black solid line) are compared to the system without an electric field (blue dashed line). The GNR has 154 valence bands and the applying of a transverse electric field opens up two finite gaps at \(k=\pi/a\) between 153 and 154, and 155 and 155 bands. A band inversion at the two finite gaps is induced by switching the direction of the electric field, which is supported by the change of parity of the wave functions for the middle four bands at \(k=\pi/a\). This can be seen by comparing (b) to (f), which are the distributions of amplitude of wave functions. As the GNR is mirror symmetric, the Zak phase of the four bands can therefore be calculated according to Eq. 4. It suggests a topological phase change by electric field if the system is filled up to \(153^{th}\) or \(155^{th}\) bands. Boundary states are expected to exist within the gap labeled in red. (c) and (g) are band structures of the square-root model with parameters chosen according to the gap width in (a) and (e). The resultant parity of the wave functions at the middle of the Brillouin zone matches that of pzc-GNR, as shown by comparing (d) to (b), and (h) to (f). Energy is with respect to the Fermi level. The green plus and orange minus signs indicate the parity of the wave function.}\label{fig2}
\end{figure}

\clearpage

\begin{figure}[h]
\centering
\includegraphics[width=0.9\textwidth]{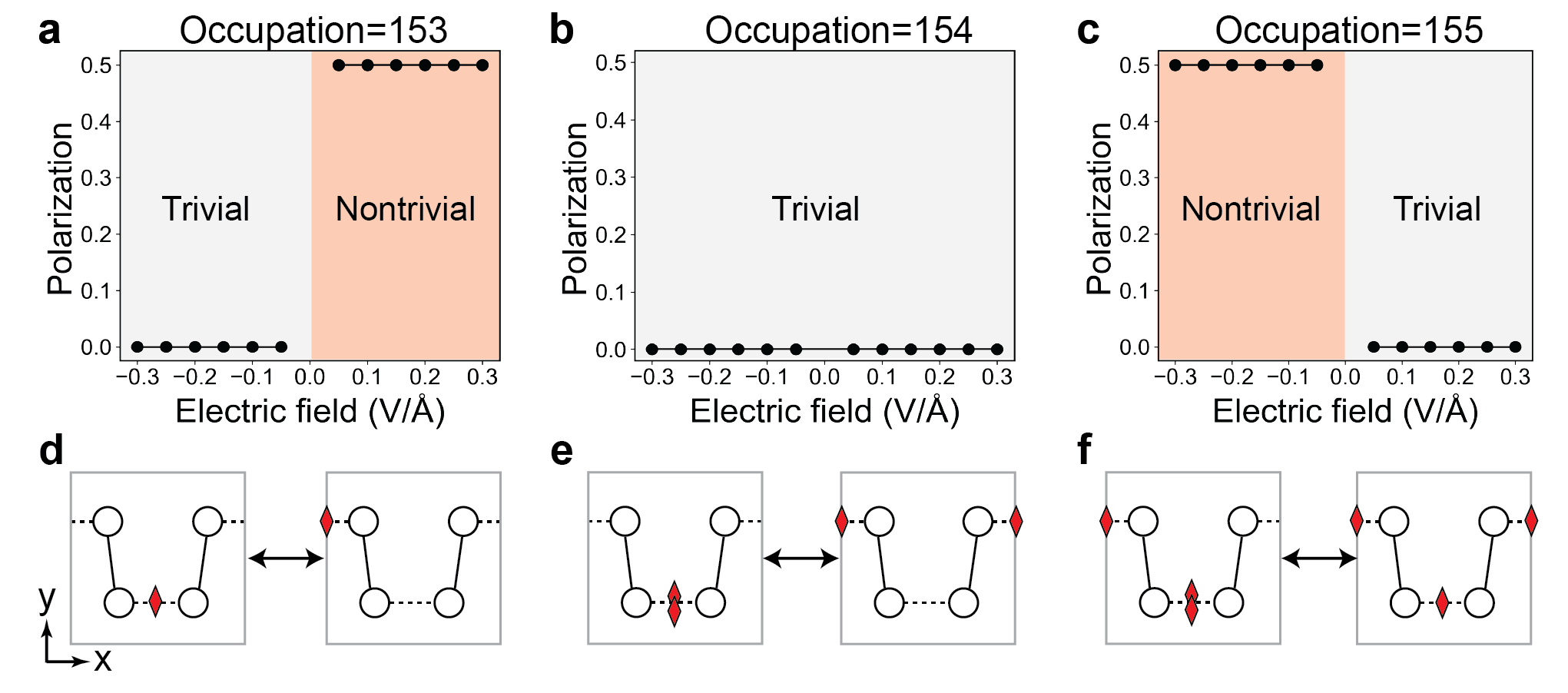}
\caption{Topological phase diagram of pzc-GNR. Polarization of pzc-GNR with (a) 153 occupied bands and (b) 154 occupied bands, and (c) 155 occupied bands. With HOMO unoccupied, or LUMO occupied, the system has topological phase transition by flipping the direction of the electric field. (d), (e), and (f) are distributions of the Wannier charge center(s) (diamond shape) of the four-band square-root Hamiltonian with one, two, and three occupied bands, respectively. Consistent with Fig. 1, x is the periodic direction and electric field is applied along the y axis. Flipping the direction of the electric field shifts the Wannier centers vertically, which inevitably changes their x coordinates. As a result, polarization (Eq. 5) with odd number of wannier centers alternates between two quantized values.}\label{fig3}
\end{figure}
 
\clearpage

\begin{figure}[h]
\centering
\includegraphics[width=0.9\textwidth]{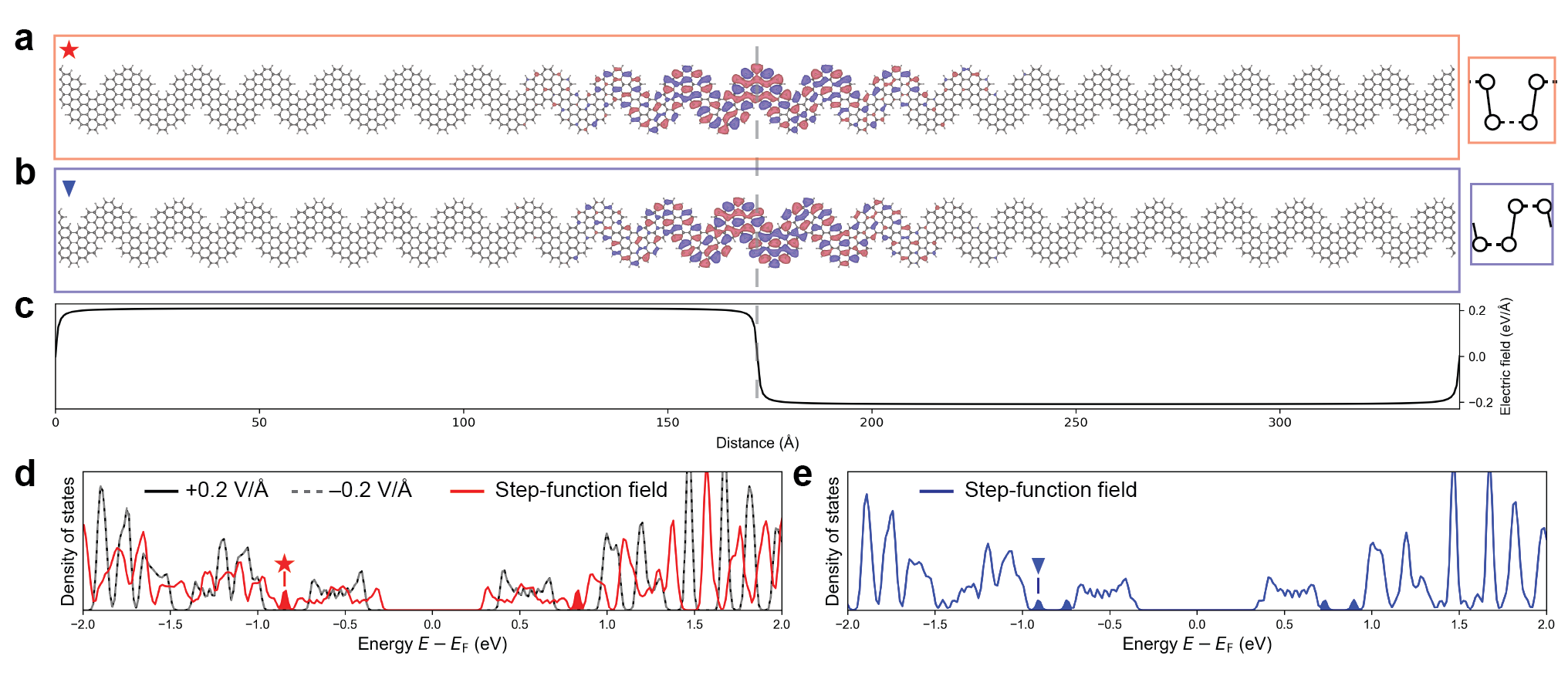}
\caption{Topological solitons. (a) A periodic supercell of pzc-GNR comprising 20 mirror-symmetric unit cells is subject to a step-function electric field, which results in a localized boundary state (soliton state) with the 0.5\% isosurfaces of the wave function plotted. (b) Same electric field is applied on a supercell consisting of 20 non-mirror-symmetric unit cells, which also yields a localized boundary state. (c) A positive and constant electric field of 0.2V/\text{\AA} is applied to the left domain of the supercell, which is transitioned to a negative one for the right domain. (d) Compared to DOS of pzc-GNR under homogeneous transverse electric field without domain walls (black and gray), DOS under a step-functioned electric field (red) shows four in-gap soliton states in total, two in each finite gap and are degenerate (peaks filled in red). (e) With another choice of domain wall configuration, the corresponding unit cell commensurate with it becomes non-mirror-symmetric. As a result, the soliton states within the finite gap can still be localized but are no longer degenerate (peaks filled in blue). Vertical dashed line indicates the center of the domain wall where the electric field changes sign. The star and triangle symbols in (d) and (e) refer to soliton states of which the wave functions are plotted in (a) and (b).}\label{fig4}
\end{figure}

\end{document}